\documentclass{aa}

\usepackage{txfonts}
\usepackage{supertabular}
\usepackage{rotating}
\usepackage{dsfont}
\usepackage{graphicx}
\usepackage{color}
\usepackage{tabularx}
\usepackage{calc}
\usepackage{array}
\usepackage[authoryear]{natbib}

\begin{document}

\titlerunning{Asymmetries in the angular distribution of the CMB}

\title{Asymmetries in the angular distribution of the cosmic microwave background}

\author{L. Santos \inst{1}  \and T. Villela \inst{2} \and C. A. Wuensche \inst{2}}

\institute{Universit\`a di Roma ``Tor Vergata'', Dipartimento di Fisica, Roma, Italy \and Instituto Nacional de Pesquisas Espaciais - INPE, Divis\~ao de Astrof\'isica, S\~ao Jos\'e dos Campos, SP, Brazil}

\offprints{L. Santos \email{ larissa.santos@roma2.infn.it}}

\date{Recieved      /Accepted}

\abstract  {Intriguing features in the angular distribution of the cosmic microwave background (CMB), such as the north-south asymmetry, were reported in the one- and three-year Wilkinson Microwave Anisotropy Probe (WMAP) data and should be studied in detail. We investigate some of these asymmetries in the CMB temperature angular distribution considering the $\Lambda$CDM model in the three, five and seven year WMAP data.}  
{We aim to analyze the four quadrants of the internal linear combination (ILC) CMB maps using three different Galactic cuts: the WMAP KQ85 mask, a $|b|<10^\circ$ Galactic cut, and the WMAP KQ85 mask $+$ $|b|<10^\circ$ Galactic cut. } 
{We used the two-point angular correlation function (TPCF) in the WMAP maps for each of their quadrants. The same procedure was performed for 1000 Monte Carlo (MC) simulations that were produced using the WMAP team  $\Lambda$CDM best-fit power spectrum. In addition, we changed the quadrupole and octopole amplitudes obtained from the $\Lambda$CDM model spectrum. We changed this to fit the quadrupole and octopole amplitudes to their observable values from the WMAP data. We repeated the analysis for the 1000 simulations of this modified $\Lambda$CDM model, hereafter M$\Lambda$CDM. } 
{Our analysis showed asymmetries between the southeastern quadrant (SEQ) and the other quadrants (southwestern quadrant (SWQ), northeastern quadrant (NEQ) and northwestern quadrant (NWQ)). Over all WMAP ILC maps, the probability for the occurrence of the SEQ-NEQ, SEQ-SWQ and SEQ-NWQ asymmetries varies from 0.1\% (SEQ-NEQ) to 8.5\% (SEQ-SWQ) using the KQ85 mask and the KQ85 mask $+$ $|b|<10^\circ$ Galactic cut, respectively. We also calculated the probabilities for the M$\Lambda$CDM using only the KQ85 mask and found no significant differences in the results. Moreover, the cold spot region located in the SEQ quadrant was covered with masks of 5,10 and 15 degrees radius and again the results remained unchanged.  Furthermore,  this analysis was repeated for random regions in the SEQ quadrant with a 15-degree mask and the SEQ quadrant still remained asymmetric with respect to the other quadrants of the CMB map.} 
{We found an excess of power in the TPCF at scales $>$ 100 degrees in the SEQ with respect to the other quadrants that is independent of the Galactic cut used. Moreover, we tested a possible relation between the Cold Spot and the SEQ excess of power and found no evidence for it. Finally,  we could not find any specific region within the SEQ that might be considered responsible for the quadrant asymmetry.}


\maketitle

\section{Introduction}

After the cosmic microwave background (CMB) discovery by A. Penzias e R. Wilson \citep{penzias1965}, several experiments were developed to characterize this radiation, leading to high-precision observations that raised the status of the CMB, which is since considered to be one of the main pillars of the $\Lambda$CDM model.  Particularly,  the CMB cosmological fluctuations first detected by the Cosmic Background Explorer (COBE) satellite data \citep{1992smoot} set up a new era in cosmology studies and paved the way to what today is called precision cosmology. However, detailed studies of the angular distribution of temperature fluctuations showed unexpected results if analyzed within the framework of the so-called cosmological concordance model ($\Lambda$CDM model). These peculiar features in the CMB angular distribution were found for the first time in the COBE data and attracted much interest since then. 

%the measurement of the CMB dipole anisotropy  - see, e.g., \citep{1983fixsen, 1983lubin, 1985lubin, 1993kogut, 1996lineweaver} - and

A quadrupole amplitude smaller than that expected according to the $\Lambda$CDM model was reported by the COBE team \citep{1992smoot} and was confirmed by all WMAP data releases \citep{2003bennett.1, 2007hinshaw,2009hinshaw,2011jarosik}. Other anomalies were found in the WMAP data that were not expected according to the $\Lambda$CDM model either, such as
the alignment between the quadrupole and octopole 
(e.g. \citep{2004bielewicz,2004schwarz,2004copi,2004deoliveiracosta,2005bielewicz,2005land,2006copi,2006abramo,2010frommert,2010gruppuso}), 
the low quadrupole and octopole amplitudes (e.g., \citep{2003mukherjee,2010ayata,2010cayon,2010cruz}), 
the north-south asymmetry 
(e.g., \citep{2004eriksen,2004hansen,2004eriksen.2,2004hansen.2,2005donoghue,2009hoftuft,2010paci,2010pietrobon,2010vielva}), 
the anomalous alignment of the CMB features toward the Ecliptic poles 
(e.g., \citep{2006wiaux,2007vielva}), 
and  the cold spot (e.g., \citep{2004vielva,2005cruz,2007cruz,2010vielva.2}). 

Recently, \citet{2011aluri} analyzed in detail the signature of parity asymmetry first  found by \citet{2010kim} in the WMAP best- fit temperature power spectrum, confirming this asymmetry on a 3-$\sigma$ level. \citet{2011aluri} also concluded  that their result is not due to residual foregrounds or to foreground cleaning. In addition, the preferred direction of the parity asymmetry coincides with the CMB kinematic dipole, showing that it may somehow be related to the quadrupole-octopole alignment \citep{2011naselsky}.

On the other hand,  \citet{2011bennett}  reviewed the anomalies reported in the CMB temperature fluctuations and claimed that they are not statistically significant and, for this reason, do not in disagree with the $\Lambda$CDM concordance model. 

Nevertheless, in this work we report an asymmetry in the WMAP temperature anisotropy data appearing in the two-point angular correlation function (TPCF) at scales above 100 degrees. In Section 2 we present  our method to prepare the MC sky map simulations to confront them with real data and describe our TPCF method. In Section 3 we show our results. Finally, in Section 4 we present the discussion, followed by the conclusions in Section 5.

\section{Method}

The results presented in this paper were derived from the analysis of three temperature ILC maps from the third, fifth and seventh year of WMAP data: WILC3 \citep{2007hinshaw}, WILC5 \citep{2008gold} and WILC7 \citep{2011jarosik}.  We computed the TPCF for the WMAP data and for the MC simulations to obtain, evaluate and finally compare the results with the $\Lambda$CDM model. The HEALPix (hierarchical equal area and isolatitude pixelization) package (synfast) \citep{2005gorski} was used to generate the MC simulations and analyze the maps. 

The simulated sky maps were generated with two different seed spectra. In the first run, we used the WMAP5 best-fit spectrum to the $\Lambda$CDM model, available at LAMBDA\footnote{http://lambda.gsfc.nasa.gov},  to generate 1000 simulations with $N_{side}=256$ (pixel diameter $ \sim 14'$). In the second run, we modified the $\Lambda$CDM model spectrum, substituting the amplitudes of the best-fit quadrupole and octopole by the values reported in the WMAP five-year data. This modified spectrum was used to generate 1000 simulations with the same resolution as the first run (the M$\Lambda$CDM model).

The  data and the simulations were analyzed by means of the TPCF. Since it is hard to analyze the CMB angular distribution because of the foreground contamination, even using multifrequency maps, we chose to use three different Galactic cuts. When the first CMB maps covering the whole sky  were released from the Relict satellite \citep{1984strukov}, the safest way  to deal with the foreground contamination was making parallel cuts above and below the Galactic plane. More recently, the masks developed by the WMAP team remove in a fair way the known point sources and the Galactic signal, but they are not perfect since the Galactic foregrounds are still not completely known. \citet {1992smoot} also warned that  the TPCF is highly affected when regions  $|b| < 10^\circ$ are included in the analyses.  Taking into account these results, we used the WMAP KQ85 mask, a Galactic cut $|b|<10^\circ$ and the WMAP KQ85 mask $+$ $|b|<10^\circ$ Galactic cut in the present work. This last cut avoids the Galactic foreground and also the known point sources. All calculations were performed for degraded maps with $N_{side}=64$. We analyzed the NWQ with 9906, 10200 and 9495 pixels for each Galactic cut.  For the SWQ, 10423, 10200 and 9795 pixels were used, for the NEQ, 10250,10176 and 9417 pixels were used, and for the SEQ 9854, 10176, 9350 pixels were used, when the Galactic foreground was removed with the WMAP KQ85 mask, the Galactic cut $|b|<10^\circ$, and the WMAP KQ85 mask $+$ $|b|<10^\circ$ Galactic cut, respectively. We also counted the pixels left in each quadrant when the new WMAP mask KQ85y7 was used to remove the Galactic foregorund, there were 9571 pixels in the NWQ, 10093 pixels in the SWQ, 9876 pixels in the NEQ, and 9217 pixels in the SEQ. 

The influence of the KQ85 mask asymmetry on the results was evaluated considering the three Galactic cuts. The same procedure was applied to the $\Lambda$CDM simulated sky maps. The M$\Lambda$CDM realizations were analyzed using only the WMAP KQ85 mask. This modified spectrum enabled us to evaluate if the observed low quadrupole and octopole values can account for the results. We did not repeat the analysis using the KQ85y7 mask for the MC simulations because the computational time for computing the TPCF is prohibitive.

\subsection{Two-point angular correlation function} \label{function}

The TPCF function measures the angular correlation of temperature fluctuations distributed in the sky and is defined as
 
\begin{equation}
c(\gamma)\equiv \langle T({\bf n_p}) T({\bf n_q})\rangle.
\end{equation}

\noindent $T({\bf n_p})$ and $T({\bf n_q})$ are the temperature fluctuations of the $p$ and $q$ pixels, respectively, and $\gamma$ is the angular distance between the two pixels.

The pixels $p$ and $q$ are defined by the coordinates ($\theta_p$, $\phi_p$) and ($\theta_q$, $\phi_q$), where $0^\circ\leq \phi \leq 360^\circ$ and $-90^\circ \leq \theta \leq 90^\circ$. It is now possible to obtain the equation for $\gamma$:

\begin{equation}
\cos\gamma = \cos\theta_p \cos\theta_q + \sin\theta_p \sin\theta_q \cos(\phi_p-\phi_q).
\end{equation}

Finally, we define an rms-like quantity, $\sigma$, to compare the TPCF computed both for WMAP data and MC simulations \citep{2006bernui}:

\begin{equation}
\label{sigma}
	\sigma = \sqrt{\frac{1}{N_{bins}}\sum_{i=1}^{N_{bins}}f_i^2}. 
\end{equation}

The $f_i$ corresponds to the TPCF  for each bin $i$. We used the number of bins $N_{bins}=90$ to quantify our results.

\section{Results}

\begin{figure}
\resizebox{\hsize}{!} {\includegraphics{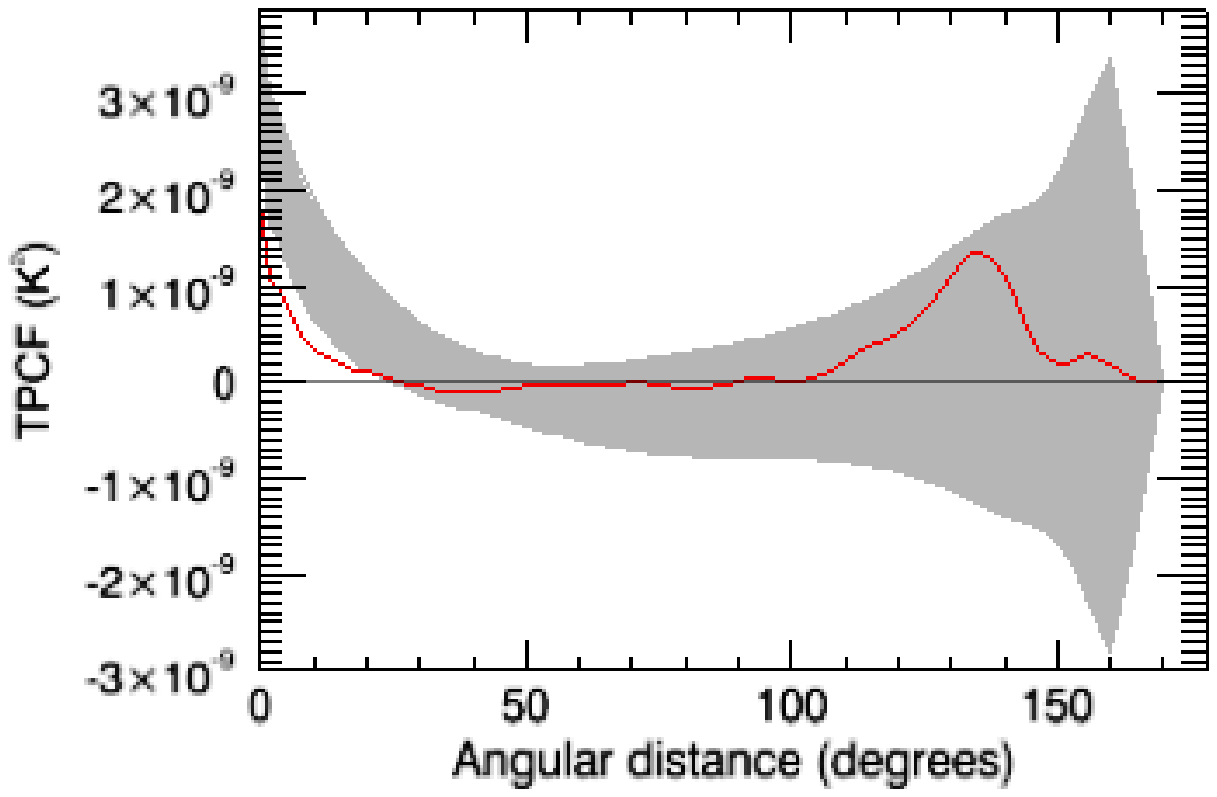}}
\resizebox{\hsize}{!} {\includegraphics{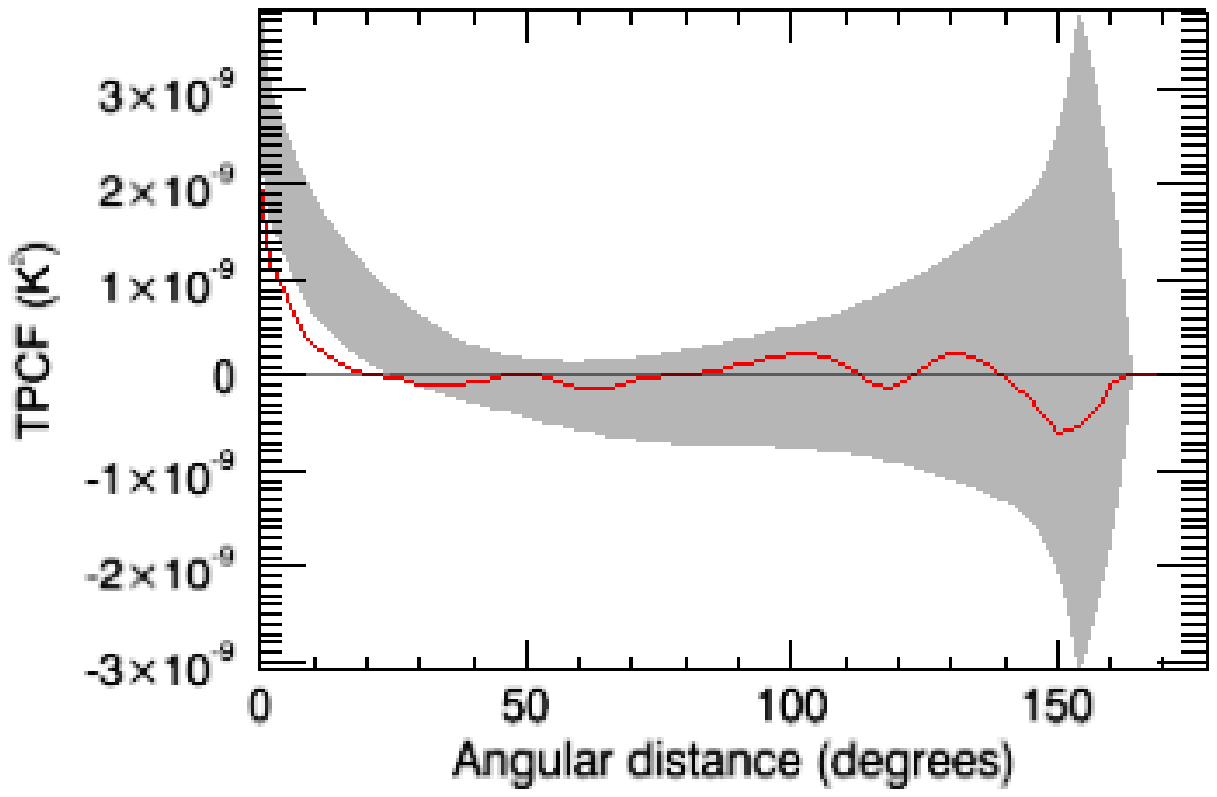}}
\caption{TPCF curves computed for the WILC7 map using the WMAP KQ85 mask, smoothed for illustration purposes.The NWQ (top) and the NEQ (bottom) appear as solid red lines. The shadow part depicts the standard deviation intervals for 1000 simulated maps produced with the $\Lambda$CDM spectrum.}
\label{TPCF-NWQ-NEQ}
\end{figure}

\begin{figure}
\resizebox{\hsize}{!} {\includegraphics{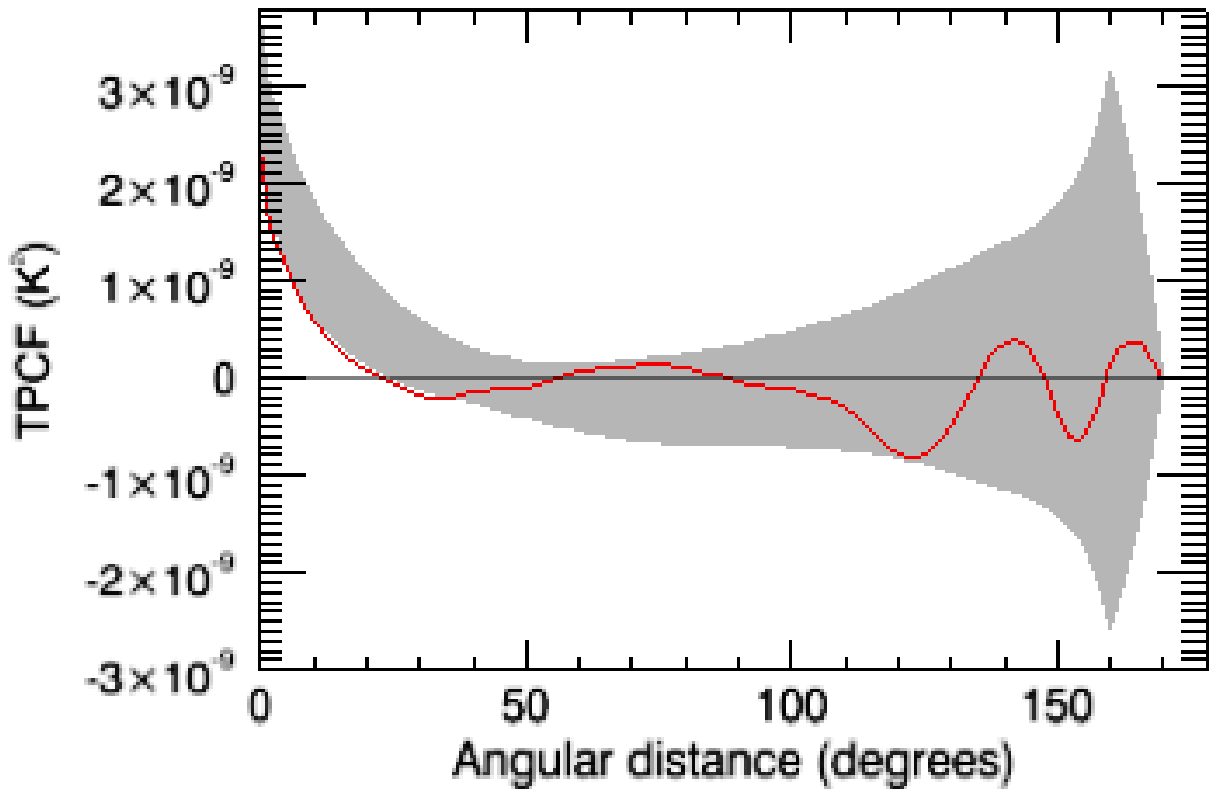}}
\resizebox{\hsize}{!} {\includegraphics{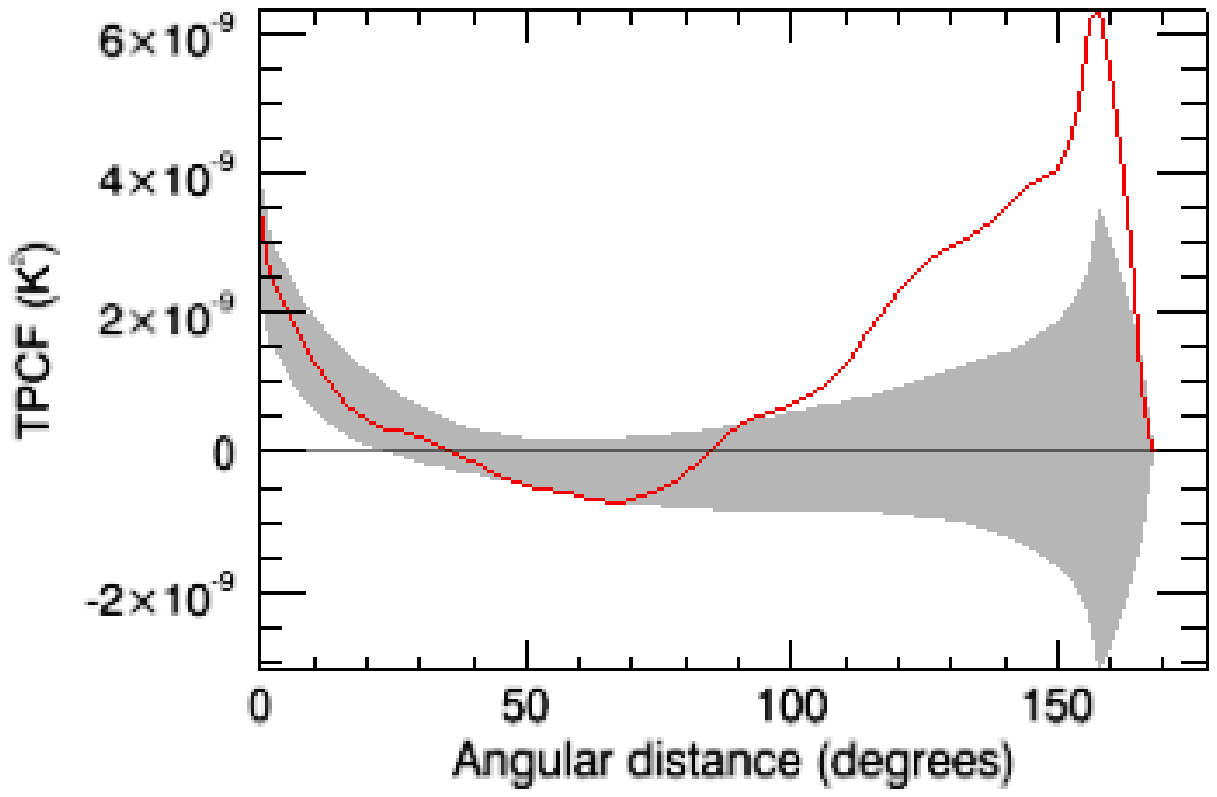}}
\caption{Same as Figure \ref{TPCF-NWQ-NEQ}, but now with the SWQ at the top and the SEQ at the bottom.}
\label{TPCF-SWQ-SEQ}
\end{figure}

\begin{figure}
\resizebox{\hsize}{!} {\includegraphics{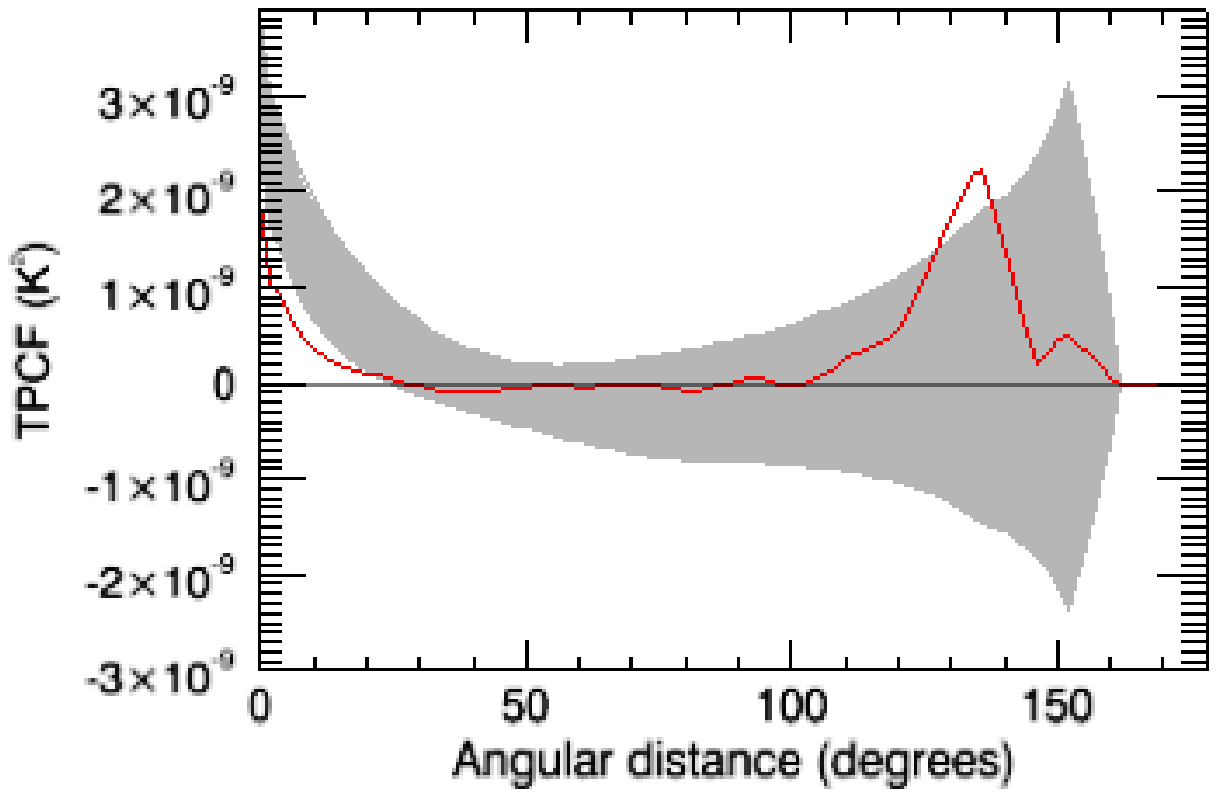}}
\resizebox{\hsize}{!} {\includegraphics{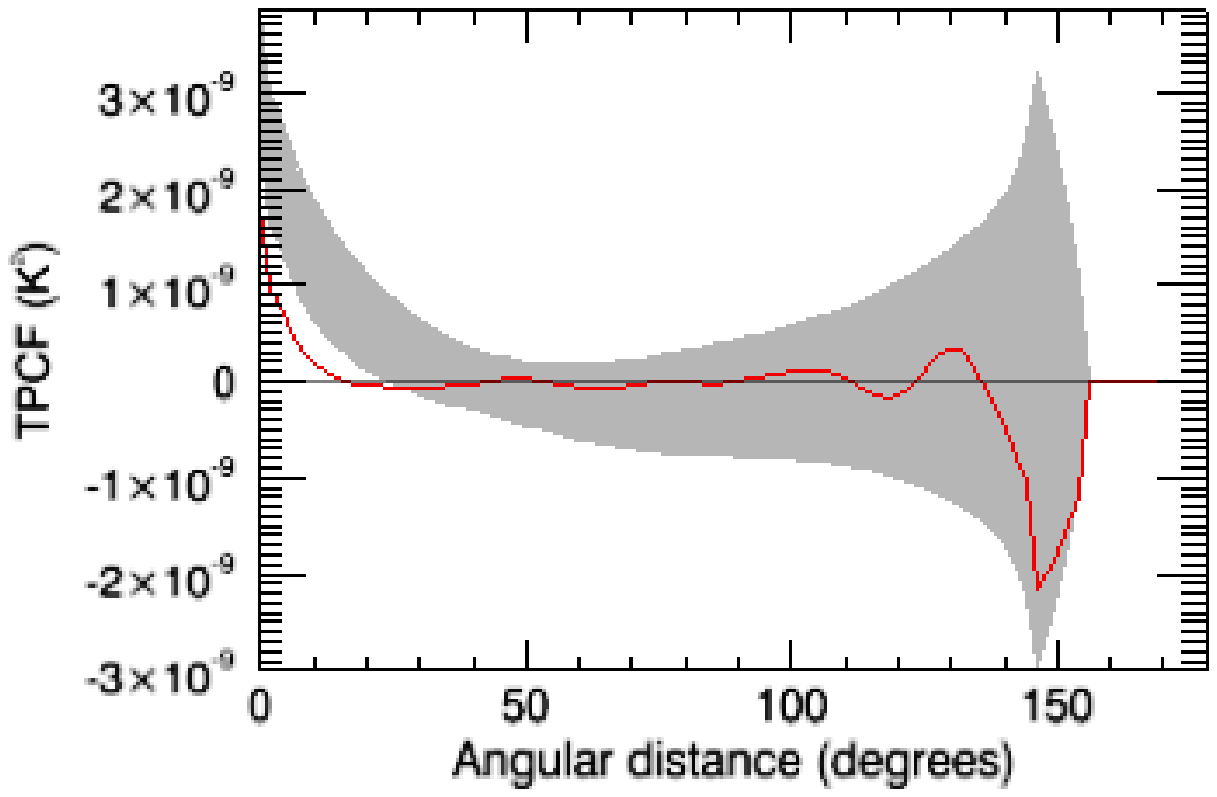}}
\caption{Same as Figure \ref{TPCF-NWQ-NEQ}, but now using the WMAP KQ85 mask $+$ $|b|<10^\circ$ Galactic cut in the temperature WMAP7 data. From  top to bottom, the curves refer to the NWQ and the NEQ, respectively.}
\label{TPCF-NWQ-NEQ-mask-cut}
\end{figure}

\begin{figure}
\resizebox{\hsize}{!} {\includegraphics{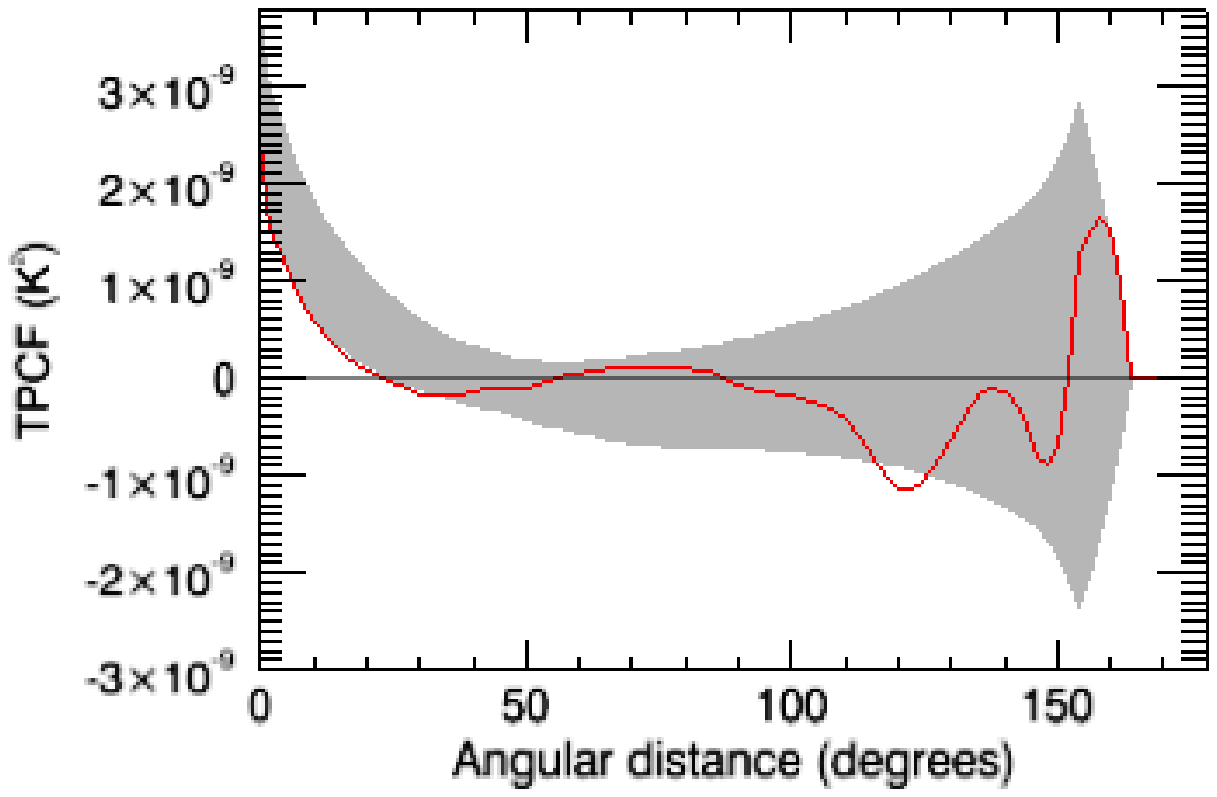}}
\resizebox{\hsize}{!} {\includegraphics{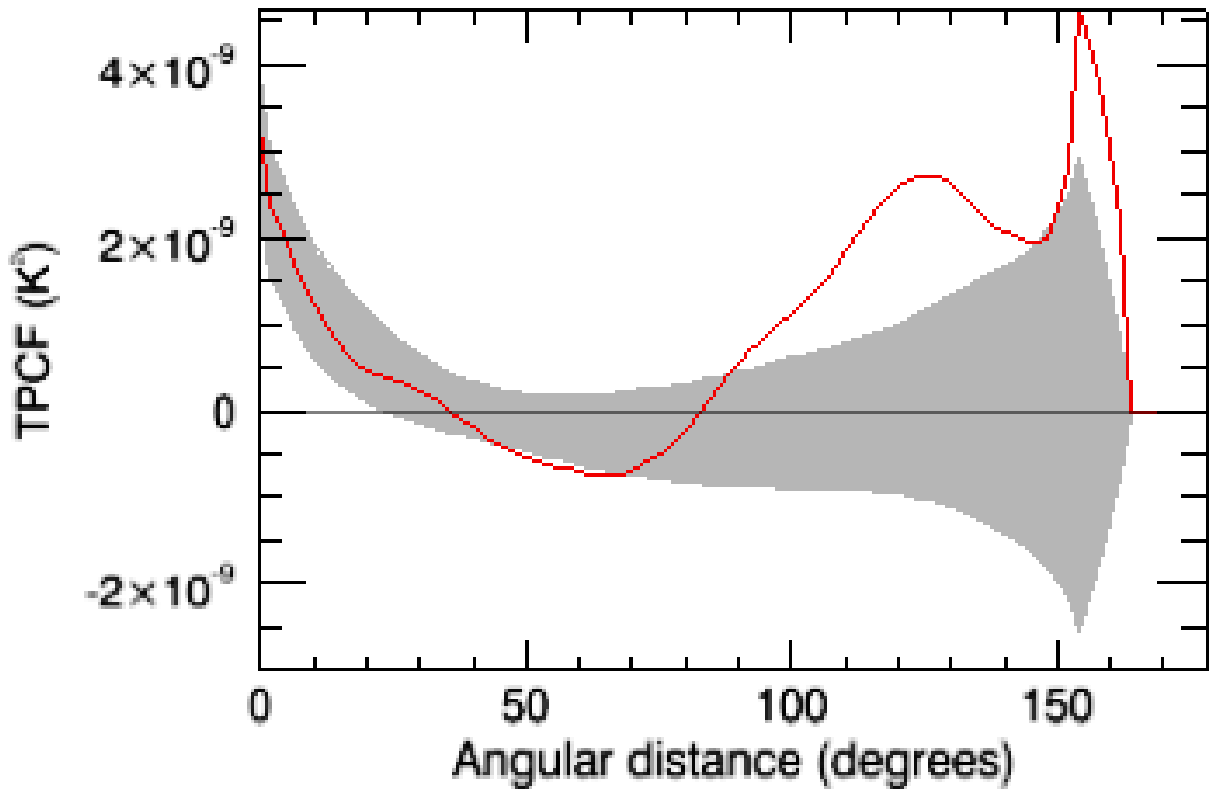}}
\caption{Same as Figure \ref{TPCF-NWQ-NEQ-mask-cut}. From top to bottom, the curves refer to the SWQ and the SEQ, respectively.}
\label{TPCF-SWQ-SEQ-mask-cut}
\end{figure}

\begin{figure}
\resizebox{\hsize}{!} {\includegraphics{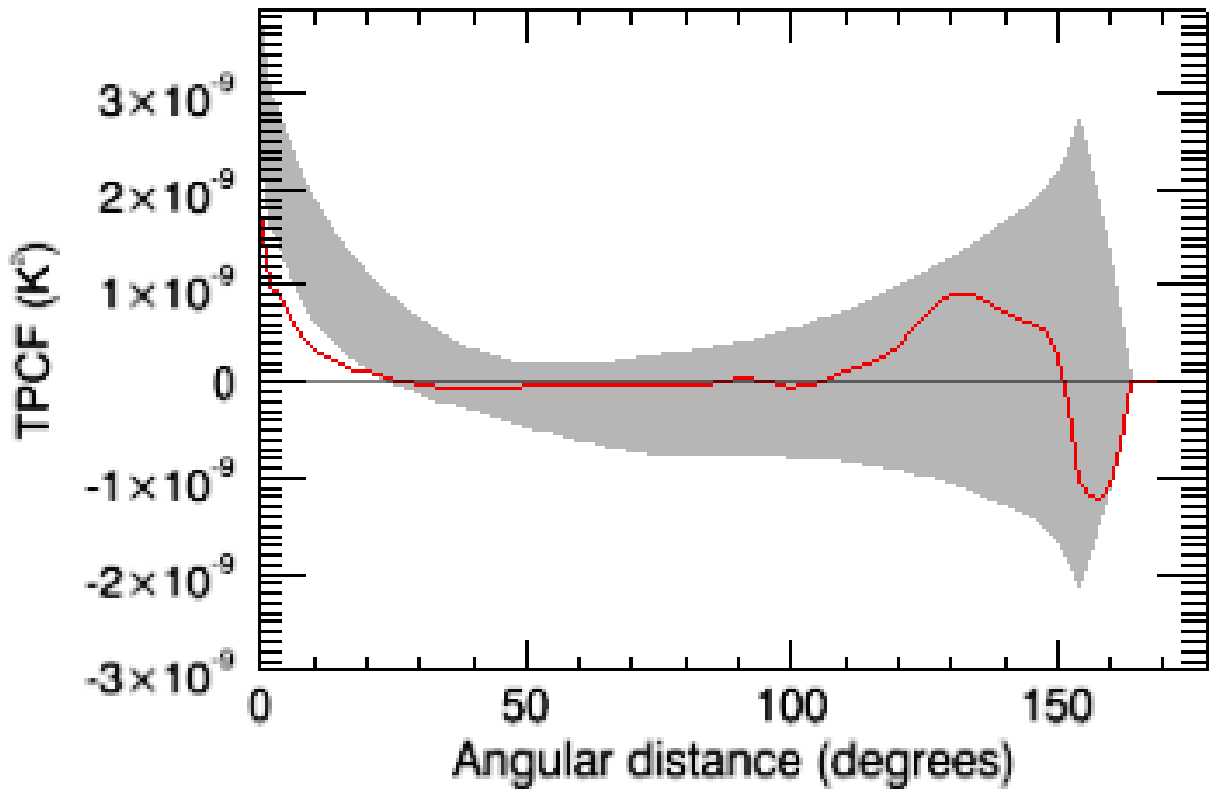}}
\resizebox{\hsize}{!} {\includegraphics{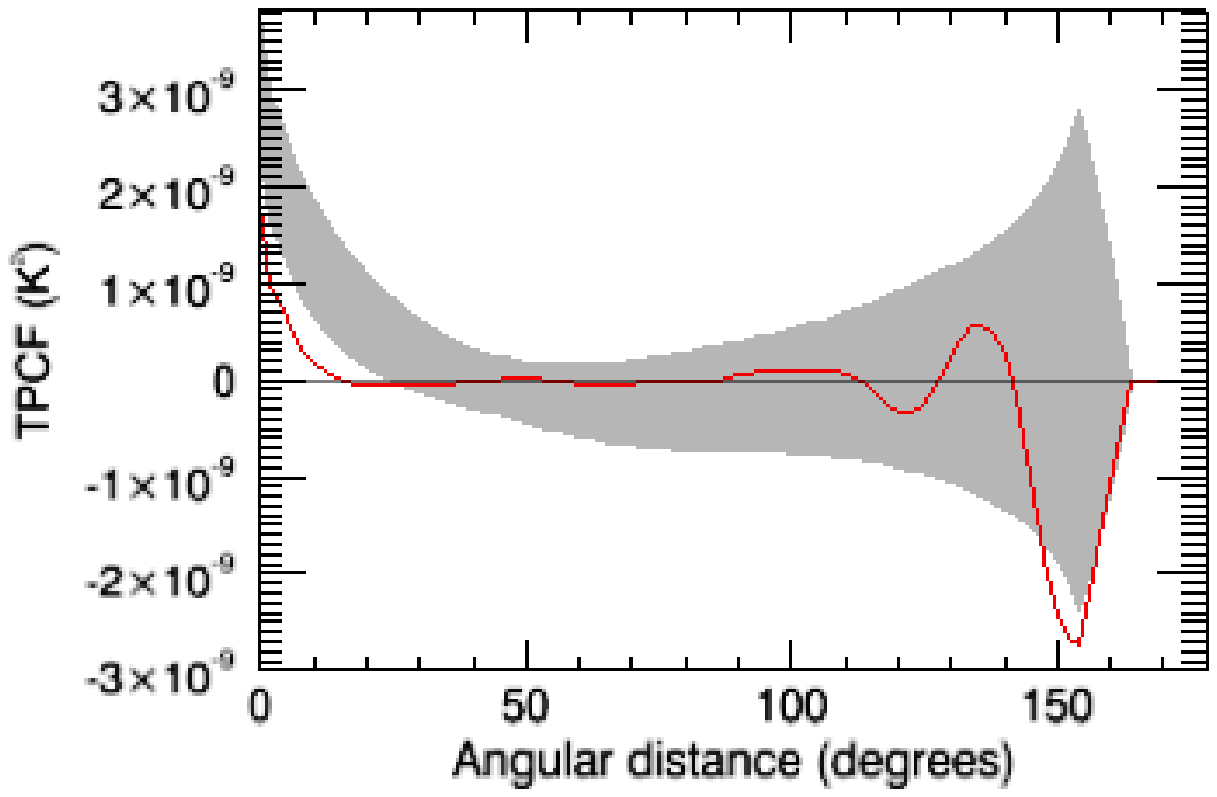}}
\caption{Same as Figures \ref{TPCF-NWQ-NEQ}-\ref{TPCF-SWQ-SEQ-mask-cut}, but now using $|b|<10^\circ$ Galactic cut, again in the temperature WILC7 map. From  top to bottom the curves refer to the NWQ and the NEQ, respectively.}
 \label{TPCF-NWQ-NEQ-cut}
\end{figure}

\begin{figure}
\resizebox{\hsize}{!} {\includegraphics{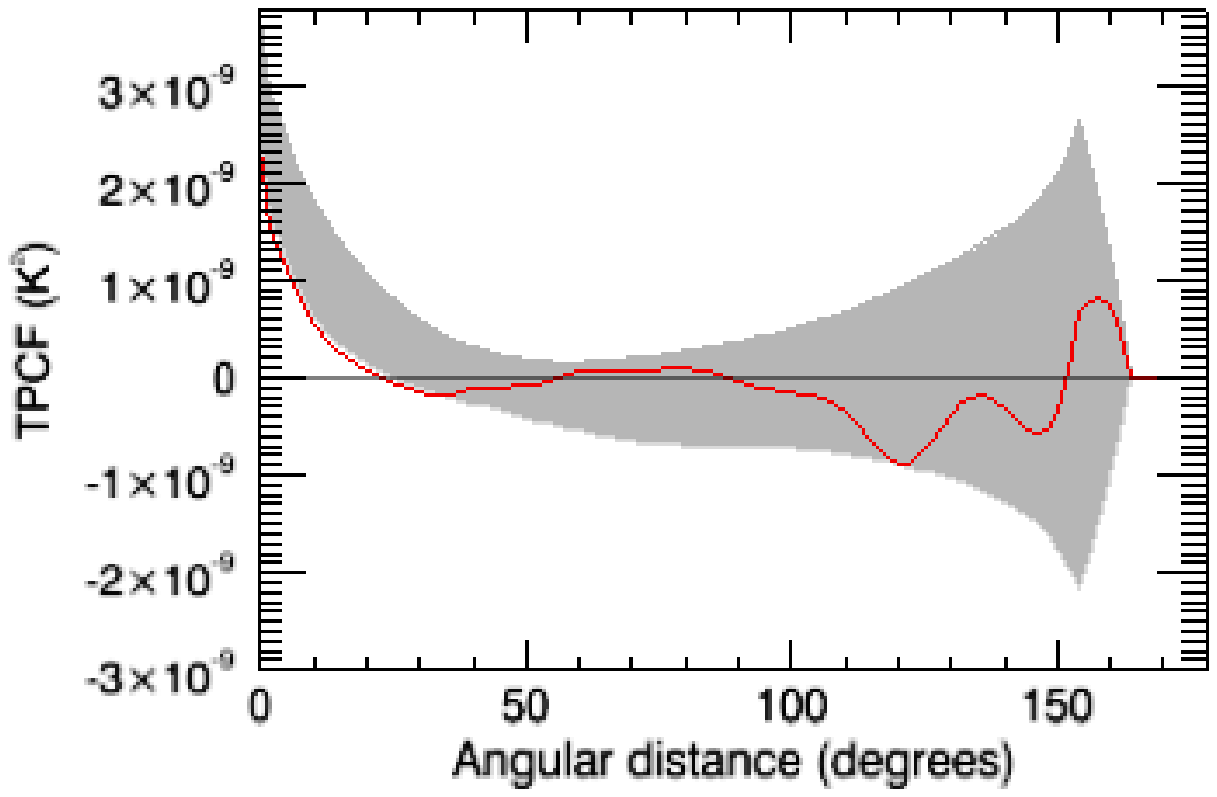}}
\resizebox{\hsize}{!} {\includegraphics{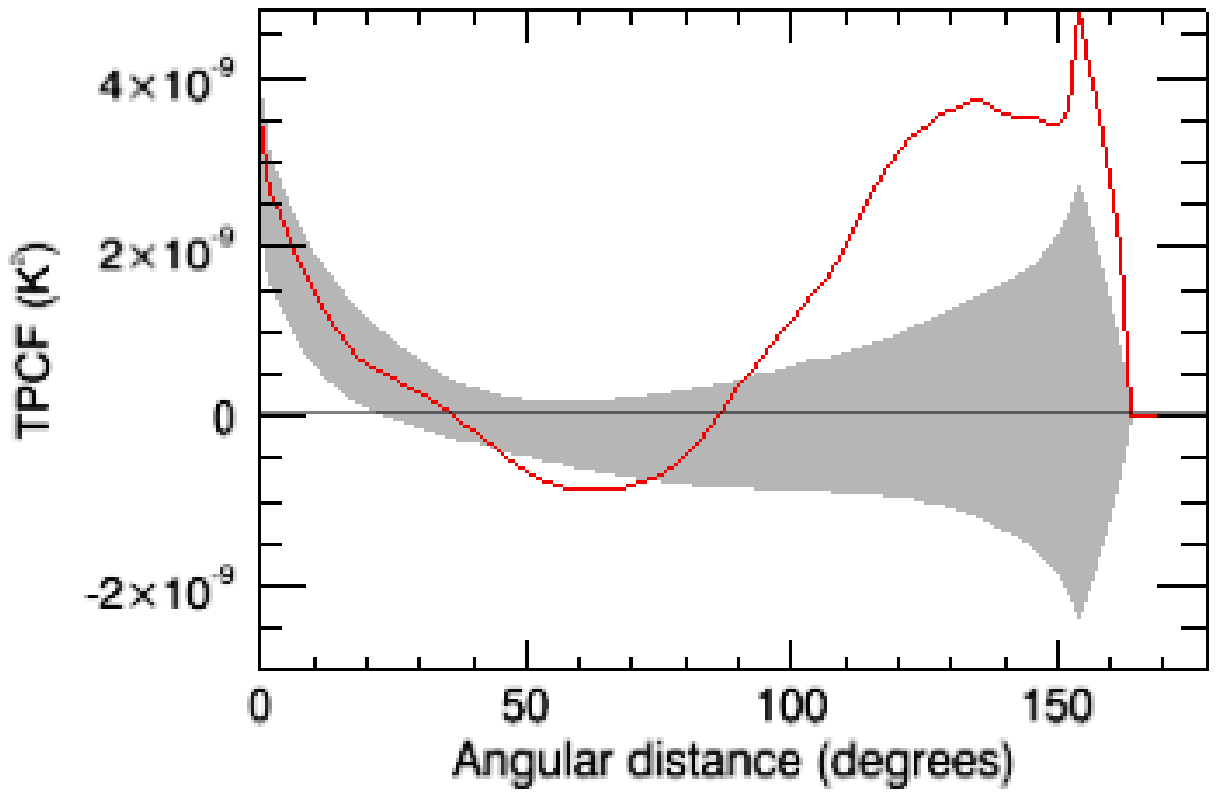}}
\caption{Same as Figure \ref{TPCF-NWQ-NEQ-cut} for the SWQ (top) and the SEQ (bottom).}
 \label{TPCF-SWQ-SEQ-cut}
\end{figure}

As mentioned before, we divided the CMB sky into quadrants  and computed the TPCF  for each quadrant, using the three different Galactic cuts: the WMAP KQ85 mask, the Galactic cut $|b|<10^\circ$ and the WMAP KQ85 mask $+$ $|b|<10^\circ$ Galactic cut. The results are shown in Figures  \ref{TPCF-NWQ-NEQ} to \ref{TPCF-SWQ-SEQ-cut}. Asymmetries between the SEQ and the other quadrants can be easily noticed in the curves. On the other hand, the TPCF curves for the NWQ, the NEQ and the SWQ  are mostly inside the gray-shadowed area, which corresponds to the MC 1$\sigma$ interval.

%\begin{figure} 
%\resizebox{\hsize}{!} {\includegraphics{./figuras/NWQ1.eps}}
%\resizebox{\hsize}{!} {\includegraphics{./figuras/SWQ1.eps}}
%\resizebox{\hsize}{!} {\includegraphics{./figuras/NEQ1.eps}}
%\resizebox{\hsize}{!} {\includegraphics{./figuras/SEQ1.eps}}
%\caption{An illustration of the NWQ, SWQ, NEQ, SEQ quadrants used in our analysis. } 
%\label{NWQ-SWQ}
%\end{figure} 

Tables \ref{tbl-prob-mask}, \ref{tbl-prob-mask-cut} and \ref{tbl-prob-cut} show the probability for these asymmetries to occur for the ILC maps, and for the MC maps derived from the $\Lambda$CDM model. Both consider the above mentioned Galactic cuts.

These tables also contain two sets of probabilities, P1 and P2. The probability P1 refers to the chance that exactly the same asymmetries found in the WMAP sky maps appear in the MC simulations:

\begin{itemize}
\item $\sigma_{SEQ}/\sigma_{NWQ} (simul.)   \geq \sigma_{SEQ}/\sigma_{NWQ} (data) $
\item$\sigma_{SEQ}/\sigma_{SWQ} (simul.)   \geq \sigma_{SEQ}/\sigma_{SWQ} (data) $
\item$\sigma_{SEQ}/\sigma_{NEQ} (simul.)   \geq \sigma_{SEQ}/\sigma_{NEQ} (data). $
\end{itemize}

The probability P2 extends the range of P1, comparing the chance that the ratio between the asymmetries found in the SEQ and in any of the three quadrants (from the MC simulations) to exceed that from the ratio between the SEQ and one given quadrant. The ** in the expressions below apply for any of the three quadrants ($** = NW ~\textrm{or}~ SW~ \textrm{or} ~NE$). 

 \begin{itemize}
\item$\sigma_{SEQ}/\sigma_{** Q} (simul.)   \geq \sigma_{SEQ}/\sigma_{NWQ} (data) $
\item$\sigma_{SEQ}/\sigma_{**Q} (simul.)   \geq \sigma_{SEQ}/\sigma_{SWQ} (data) $
\item$\sigma_{SEQ}/\sigma_{**Q} (simul.)   \geq \sigma_{SEQ}/\sigma_{NEQ} (data). $
\end{itemize}

\subsection{The cold spot}

Because the asymmetry was coincidentally found in the same quadrant as the so-called cold spot (see \citep{2007cruz,2010vielva}), we tested if they are possibly correlated.  We masked the cold spot region which is centered on $\phi = 209^\circ$ and $\theta=141^\circ$, and calculated the TPCF for the SEQ using also the KQ85 mask.  The cold spot mask  radii were $5^\circ$, $10^\circ$ and $15^\circ$.

%\begin{figure}
%\resizebox{\hsize}{!}{\includegraphics{./figuras/cold-spot-QID-5-mask85.eps}}
%\resizebox{\hsize}{!}{\includegraphics{./figuras/cold-spot-10-QID-mask85.eps}}
%\resizebox{\hsize}{!}{\includegraphics{./figuras/cold-spot-QID-15-mask85.eps}}
%\caption{ First it is shown the SEQ + the KQ85 mask and a mask used to cover the Cold Spot region with a radius of 5, 10 and 15 degrees.}
 %\label{TPCF-cold-spot} 
%\end{figure}

\begin{figure}
\resizebox{\hsize}{!}{\includegraphics{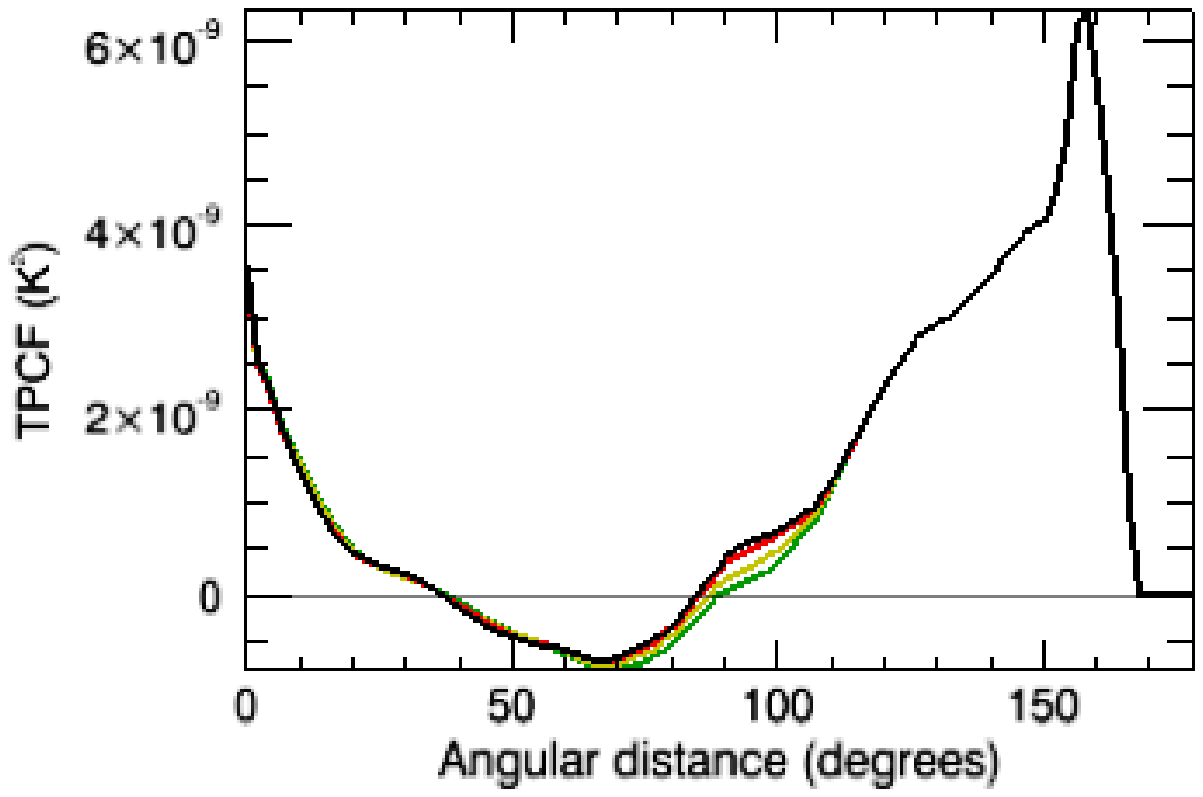}}
\resizebox{\hsize}{!}{\includegraphics{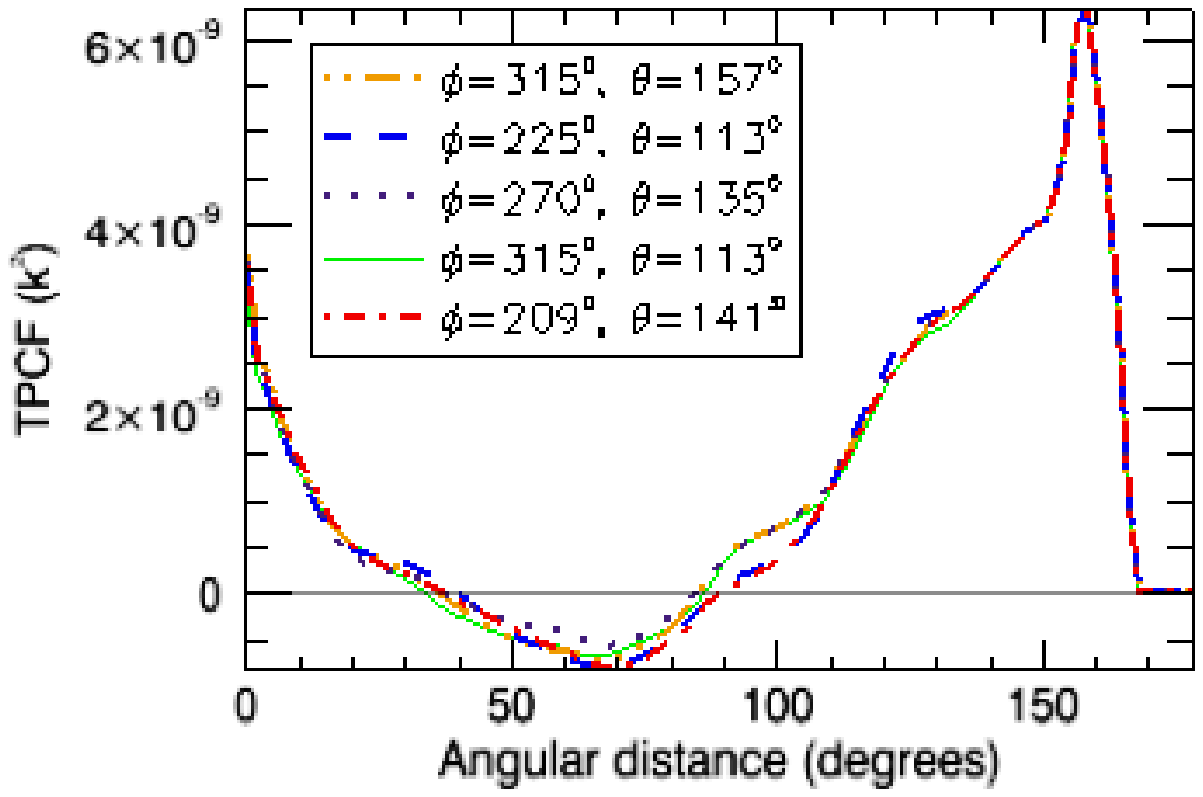}}
\caption{Top: Comparison between the SEQ quadrant for the TPCF using the KQ85 mask + these circular masks centered on the cold spot can be seen with a radius of 5 degrees (red line), 10 degrees (yellow line) and 15 degrees (green line).  The black line refers to the function without masking the Cold Spot. Bottom: TPCF for the WMAP7 map using KQ85 + circular masks centered on the specified angles with a radius of 15 degrees each. The red dashed-dot curve refers to the Cold Spot region.}
 \label{varredura} 
\end{figure}

Finally, to look for a specific region in the SEQ that could account for the asymmetry, we scanned this quadrant, masking four different regions in addition to the cold spot region. The masks were centered on coordinates chosen randomly on ($\phi = 315^\circ, \theta=157^\circ$), ($\phi = 225^\circ, \theta=113^\circ$), ($\phi = 270^\circ,\theta=135^\circ$), ($\phi = 315^\circ, \theta=113^\circ$), all of them with a radius of $ 15^\circ$. The TPCF was computed using a mask in each position at a time. The results are shown in Figure \ref{varredura} (bottom).

\section{Discussion}

Our analyses indicate that, even though the excess of power in the SEQ is independent of the chosen Galactic cut, the largest asymmetry between the SEQ and the other quadrants depends on the Galactic cut and also on which ILC map is used. Considering the analysis performed only with the KQ85 mask, the biggest asymmetry was found between the SEQ and the NEQ for all three ILC maps, the largest corresponding to the WILC5 map. Moreover, when only the KQ85 mask was used, the asymmetries between the SEQ and the other three quadrants for the other Galactic cuts are smaller. However, we found that the chance of having any asymmetry is quite low. Table 4 shows that most of the sigma ratios between the SEQ and the other three quadrants obtained from the TPCF computed from the data (see Eq. \ref{sigma}) are outside the standard deviation interval of the sigma ratios obtained from the TPCF computed for the MC simulations.

%From the analysis done we could verify that, even though the excess of power in the SEQ is independent of the Galactic cut used,  the largest asymmetry between the SEQ and the other quadrants depends on the Galactic cut and also on which  ILC map is  used.  Considering the analysis done only with the KQ85 mask, the biggest asymmetry was found between the SEQ and the NEQ for all three ILC maps, being larger for the WILC5 map. Moreover, the asymmetries between the SEQ and the other three quadrants for the other Galactic cuts  are not as large as when only the KQ85 mask is used. However, we found that the chance of having any asymmetry is low and most of the sigma ratios between the SEQ and the other three quadrants obtained from the TPCF applied to the data (see Eq. \ref{sigma}) are outside the standard deviation interval of the sigma ratios obtained from the TPCF applied to the MC simulations, as shown in Table \ref{tbl-sigma}. 

Indeed, most of the sigma ratios between the SEQ and any other quadrant obtained from the TPCF computed for the simulated sky maps average one, as expected, because we generated realizations with randomly distributed temperature fluctuations in the sky.  

\begin{figure}
\resizebox{\hsize}{!}{\includegraphics{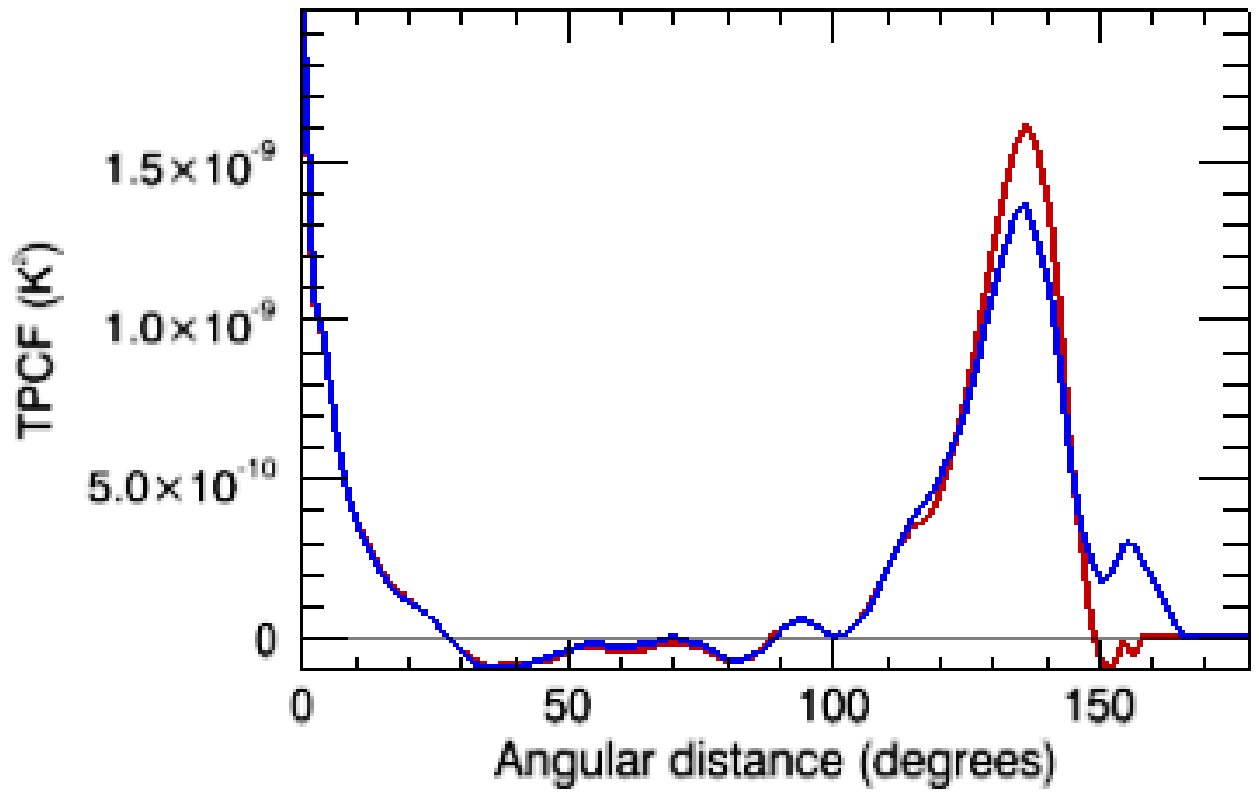}}
\resizebox{\hsize}{!}{\includegraphics{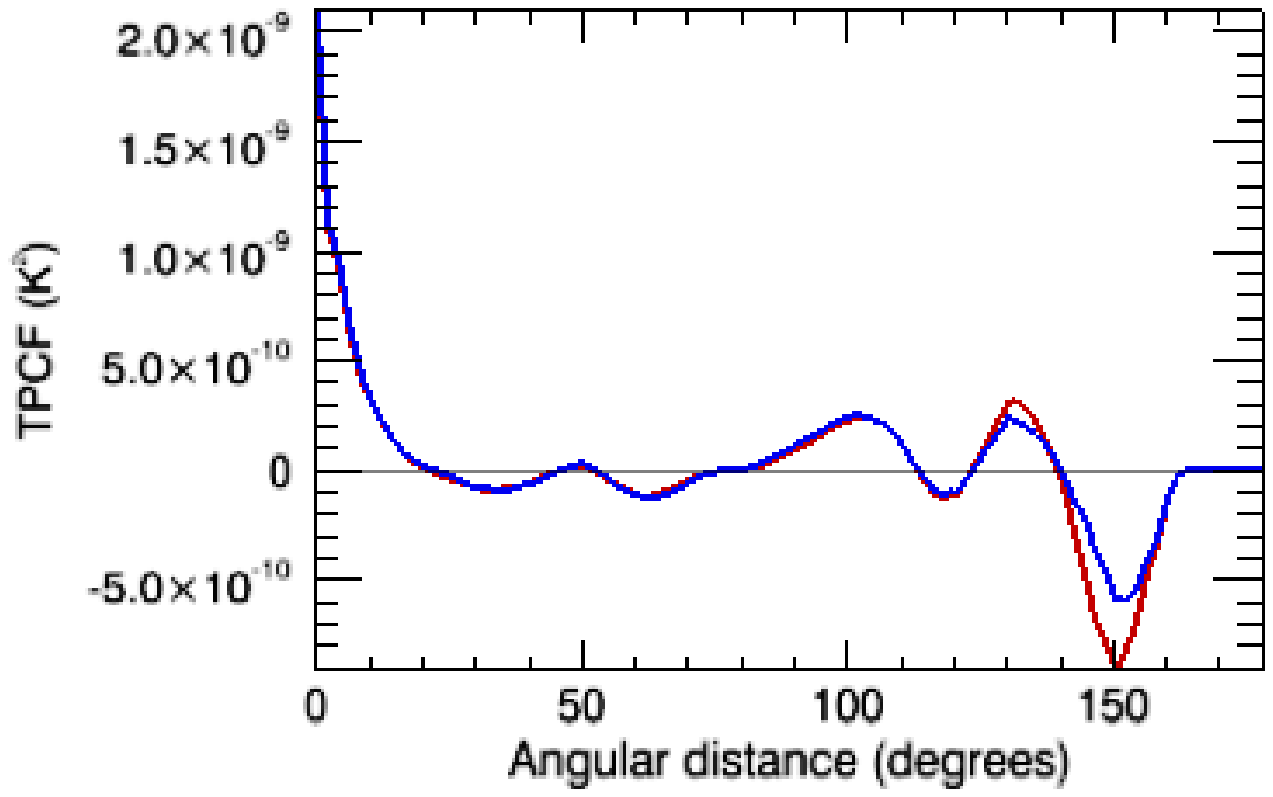}}
\caption{Comparison between NWQ (top) and NEQ (bottom)  for the TPCF using the KQ85 mask (solid blue line) and the KQ85y7 (solid red line).} 
 \label{TPCF-mask-masknova1}
\end{figure}

\begin{figure}
\resizebox{\hsize}{!}{\includegraphics{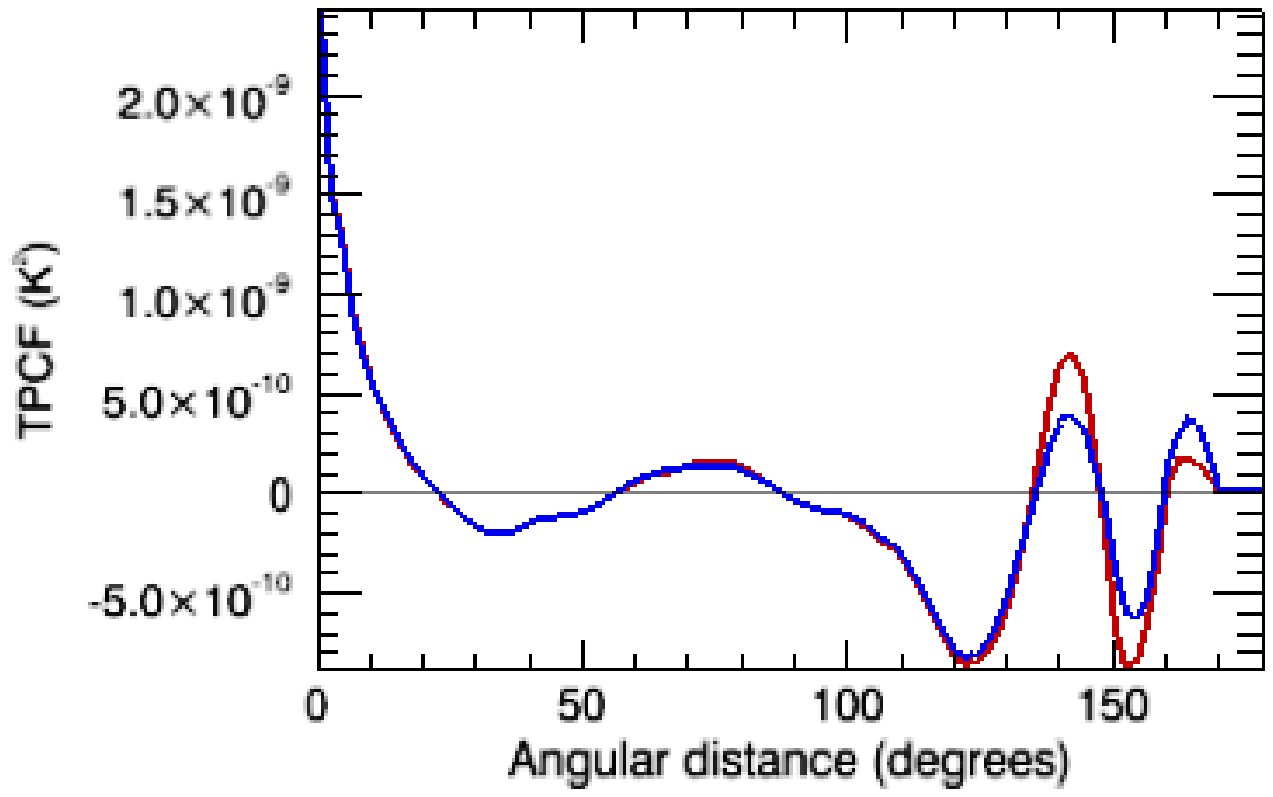}}
\resizebox{\hsize}{!}{\includegraphics{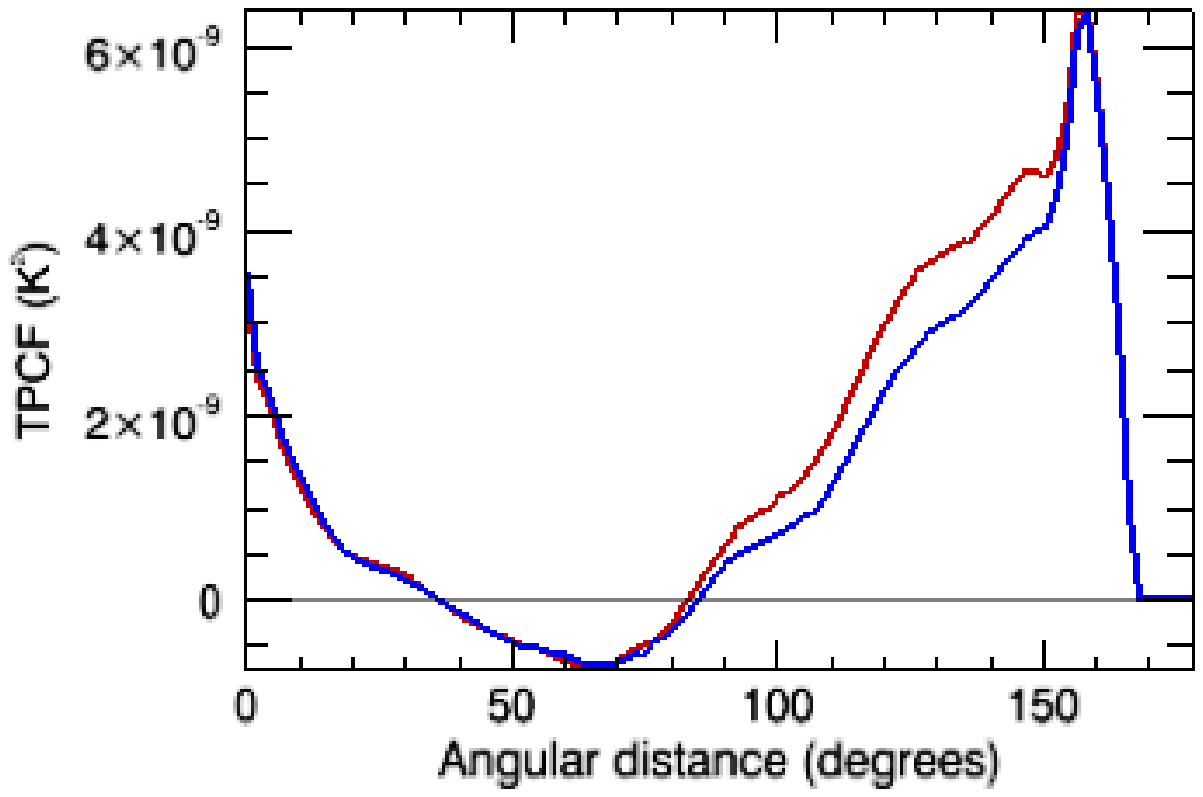}}
\caption{Same as Figure  \ref{TPCF-mask-masknova1} for the SWQ (top) and the SEQ (bottom). } \label{TPCF-mask-masknova2}
\end{figure}

Finally, it is possible to see that the excess of power in the SEQ is independent of the chosen Galactic cut. However, the chosen Galactic cut influences the size of this excess in comparison to the other quadrants. Furthermore, using the asymmetric mask does not change our results drastically. We also notice this by analyzing the effect of the new mask provided for the WMAP sseven-year data release (KQ85y7) in TPCF. There is no clear difference in the results of the TPCF for each quadrant using the WILC7 + KQ85 and WILC7 + KQ85y7 data sets, as can be seen from Figures \ref{TPCF-mask-masknova1} and \ref{TPCF-mask-masknova2}. The values found for the sigma ratios between the SEQ and the other quadrants in this last case were $\sigma_{SEQ}/\sigma_{NWQ} = 4.6$, $\sigma_{SEQ}/\sigma_{SWQ}= 4.7 $ and $\sigma_{SEQ}/\sigma_{NEQ}= 6.8$ (see Table \ref{tbl-prob-mask} for comparison). 

Furthermore, using the M$\Lambda$CDM, which includes the WMAP best-fit values for the quadrupole and octopole, instead of the $\Lambda$CDM to generate the MC simulations  does not noticeably change neither P1 nor P2, as can be seen in  Table \ref{tbl-prob-mask-mod}. Based upon these results, we are confident to state that  the low values for the quadrupole and octopole in the data are not responsible for this quadrant asymmetry.

Finally, no evidence for a relationship between the SEQ asymmetry and the cold spot was found.  The excess of power in the SEQ was most evident above the angular distance of 100 degrees. By masking the cold spot, we were able to identify small differences in the behavior of the TPCF, all of them remaining within the angular distance of 100 degrees. The same result was found for the other randomly chosen regions in the SEQ quadrant and  no evidence was found that can relate the reported asymmetry to any of these chosen regions in the SEQ.

\section{Conclusion}

We found a significant asymmetry between the SEQ and the other quadrants by considering the temperature WMAP ILC maps. We calculated the probability of occurrence for this asymmetry using MC simulations,  and 1 out of 1000 simulations for the $\Lambda$CDM model corresponded to the SEQ-NEQ asymmetry found in the WILC7 using the KQ85 mask. We also showed that different Galactic cuts do not influence the result in a significant way, leading us to believe that this effect is not caused by the asymmetric mask. Moreover, the use of KQ85y7 preserves the same asymmetries (SEQ-NEQ, SEQ-SWQ and SEQ-NWQ), as expected. Considering all Galactic cuts and maps used, the highest probability of having an asymmetry  is 8.5\%, for the WILC5 with the KQ85 mask + data clipping for $|b|<10^\circ$.

%We found a significant asymmetry between the SEQ and the other quadrants considering the temperature WMAP ILC maps. We calculated the probability of  this asymmetry occurrence by analyzing MC simulations. It was found that 1 out of 1000 simulations for the $\Lambda$CDM model corresponded to the SEQ-NEQ asymmetry found in the WILC7 using the KQ85 mask. It was also shown that different Galactic cuts do not influence the result in a significative way, leading us to infer that this effect is not due to the asymmetric mask. Moreover, the use of KQ85y7 maintain the same asymmetries (SEQ-NEQ, SEQ-SWQ and SEQ-NWQ) as expected. Considering all the Galactic cuts and maps we used, the highest probability of having an asymmetry was found for the WILC5 using the KQ85 mask $+$ $|b|<10^\circ$ is 8.5\%. 

The possibility that the asymmetries  described in this work were related to the reported lack of power in the quadrupole and octopole was tested. We constructed simulations based on a modified $\Lambda$CDM model, adjusting the amplitudes of the quadrupole and the octopole to their observational WMAP values. No explicit relation between the quadrant asymmetries and the low amplitude of the first two non-zero multipoles was found.

Furthermore, we found no evidence of a relationship between the cold spot region and the SEQ excess of power, as pointed out by \cite{2009bernui}, who suggested that the Cold Spot is responsible for 60\% of the Southern Hemisphere power. Masking this region and some other regions in the SEQ does not change the TPCF  in a significant way, leading us to conclude that the asymmetries between the SEQ and the other quadrants are not related to any specific region in the SEQ.

We conclude that the excess of power found in the SEQ is likely related to the north-south asymmetry, in which the South Hemisphere presents more power than the Northern one (see \citep{2004eriksen,2004hansen}  and \citep{2010paci,2010pietrobon,2010vielva} for recent discussions on the topic). Our results support the claims that there is indeed a north-south asymmetry and show that the excess of power occurs in the SEQ.  Additional investigation is needed to find a better explanation for the north-south asymmetry.

An explanation for these asymmetries is still missing. They could be primordial or caused by residual foregrounds or systematic effects. The upcoming CMB data from the Planck satellite when analyzed with more accurate foreground removal techniques will enable us to study in more detail the CMB anomalies reported in the literature.

Finally,  in addition to the SEQ excess of power, we can notice a lack of correlation in the SWQ, NWQ and NEQ in the TPCF in scales between 20 and 100 degrees for all Galactic cuts in the present work (see Figures \ref{TPCF-NWQ-NEQ}-\ref{TPCF-SWQ-SEQ-cut}).  A lack of correlation in the TPCF was already reported by \cite{2007copi} in scales above 60 degrees for a full sky analysis using the Kp0 mask in the first and third year of WMAP data.

We would like to thank Paolo Cabella for useful discussions. We also acknowledge the use of HEALPix packages, of the Legacy Archive for Microwave Background Data analysis (LAMBDA). L. Santos thanks CAPES-Brazil for financial support. T. Villela acknowledges CNPq support through grant 308113/2010-1. C. A. Wuensche acknowledges CNPq grant 308202/2010-4.

\clearpage
\onecolumn

\begin{table}
\begin{center}
\caption{Calculated probabilities of finding the same asymmetries as in WMAP data in the MC simulations using the WMAP KQ85 mask and considering the $\Lambda$CDM model.}
\label{tbl-prob-mask} 
\begin{tabular}{ccrcrcr}
\hline\hline
Map & $\sigma_{SEQ}/\sigma_{NWQ}$ & P1 \tablefootmark{a} & $\sigma_{SEQ}/\sigma_{SWQ}$ & P1 & $\sigma_{SEQ}/\sigma_{NEQ}$ & P1 \\
 \hline
 WILC7 &4.6  &0.5\% &4.6 &0.5\% &6.7 &0.1\%\\ 
 WILC5 &4.2 &0.9\% &4.1 &0.9\% &7.0 &0.1\%\\ 
 WILC3 &3.9 &1.3\% &4.1 &0.9\% &6.8 &0.1\%\\

\hline\hline
 Map & $\sigma_{SEQ}/\sigma_{NWQ}$ & P2 \tablefootmark{b} & $\sigma_{SEQ}/\sigma_{SWQ}$ & P2& $\sigma_{SEQ}/\sigma_{NEQ}$ & P2 \\
 \hline
 WILC7 &4.6  &1.7\% &4.6 &1.7\% &6.7 &0.2\%\\ 
 WILC5 &4.2 &2.6\% &4.1 &2.8\% &7.0 &0.1\%\\ 
 WILC3 &3.9 &3.5\% &4.1 &2.8\% &6.8 &0.1\%\\
  \hline
\end{tabular}
\end{center}
\tablefoottext{a}{Probability of finding, in this case, the asymmetry between the SEQ and the NWQ quadrants in the simulations.}
\tablefoottext{b}{Probability of finding the asymmetry between the SEQ quadrant  and  any other quadrant in the simulations.}
\end{table}

\begin{table}
\begin{center}
\caption{Calculated probabilities of finding the same asymmetries as in WMAP data in the MC simulations using the WMAP KQ85 mask  $+$ $|b|<10^\circ$ Galactic cut  and considering the $\Lambda$CDM model.}
\label{tbl-prob-mask-cut} 
\begin{tabular}{ccrcrcr}
\hline\hline
Map & $\sigma_{SEQ}/\sigma_{NWQ}$ & P1 & $\sigma_{SEQ}/\sigma_{SWQ}$ & P1 & $\sigma_{SEQ}/\sigma_{NEQ}$ & P1 \\
\hline
 WILC7 &2.7  &4.9\% &2.3 &7.4\% &2.5 &6.0\%\\ 
 WILC5 &2.3 &7.4\% &2.2 &8.5\% &3.2 &2.7\%\\ 
 WILC3 &2.4 &6.5\% &2.4 &6.5\% &3.0 &3.1\%\\
 
 \hline\hline
 Map & $\sigma_{SEQ}/\sigma_{NWQ}$ & P2 & $\sigma_{SEQ}/\sigma_{SWQ}$ & P2 & $\sigma_{SEQ}/\sigma_{NEQ}$ & P2\\
 \hline
 WILC7 &2.7 &10.1\% &2.3&16.0\% &2.5 &12.8\%\\ 
 WILC5 &2.3 &16.0\% &2.2 &18.3\% &3.2 &5.4\%\\ 
 WILC3 &2.4 &14.0\% &2.4 &14.0\% &3.0 &6.5\%\\
 \hline
\end{tabular}
\end{center}
\end{table}

\begin{table}
\begin{center}
\caption{Calculated probabilities of finding the same asymmetries as in WMAP data in the MC simulations using the $|b|<10^\circ$ Galactic cut and considering the $\Lambda$CDM model.}
\label{tbl-prob-cut} 
\begin{tabular}{ccrcrcr}
\hline\hline
Map & $\sigma_{SEQ}/\sigma_{NWQ}$ & P1 & $\sigma_{SEQ}/\sigma_{SWQ}$ & P1 & $\sigma_{SEQ}/\sigma_{NEQ}$ & P1 \\
\hline
 WILC7 &3.7  &1.6\% &4.0 &1.2\% &2.8 &4.8\%\\ 
 WILC5 &3.6 &1.8\% &3.6 &1.8\% &3.4 &2.5\%\\ 
 WILC3 &3.4 &2.5\% &3.6 &1.8\% &3.1 &3.2\%\\
 
  \hline\hline
 Map & $\sigma_{SEQ}/\sigma_{NWQ}$ & P2 & $\sigma_{SEQ}/\sigma_{SWQ}$ & P2& $\sigma_{SEQ}/\sigma_{NEQ}$ & P2 \\
 \hline
 WILC7 &3.7 &3.2\% &4.0 &2.8\% &2.8 &8.8\%\\ 
 WILC5 &3.6 &3.4\% &3.6 &3.4\% &3.4 &4.1\%\\ 
 WILC3 &3.4 &4.1\% &3.6 &3.4\% &3.1 &5.8\%\\
 \hline
\end{tabular}
\end{center}
\end{table}

\begin{table}
\begin{center}
\caption{Sigma ratios for the simulated CMB maps considering the $\Lambda$CDM model and the three Galactic cuts.\label{tbl-sigma} }
\begin{tabular}{cccc}
\hline\hline
Galactic cut &$\sigma_{SEQ}/\sigma_{NWQ}$ &$\sigma_{SEQ}/\sigma_{SWQ}$ &$\sigma_{SEQ}/\sigma_{NEQ}$\\
\hline
KQ85                     &$1.0^{+1.1}_{-0.5}$ &$1.1^{+1.2}_{-0.6}$ &$0.9^{+1.0}_{-0.5}$\\ 
KQ85  $+$ $|b|<10^\circ$ &$1.0^{+1.0}_{-0.5}$ &$1.0^{+1.0}_{-0.5}$ &$1.0^{+1.0}_{-0.5}$\\ 
$|b|<10^\circ$           &$1.0^{+1.0}_{-0.5}$ &$1.0^{+1.0}_{-0.5}$ &$1.0^{+1.0}_{-0.5}$\\

\hline

\end{tabular}
\end{center}
\end{table}

\begin{table}
\begin{center}
\caption{Calculated probabilities of finding the same asymmetries as in WMAP data in the MC simulations using the WMAP KQ85 mask and considering the M$\Lambda$CDM model.\label{tbl-prob-mask-mod} }
\begin{tabular}{ccrcrcr}
\hline\hline
Map & $\sigma_{SEQ}/\sigma_{NWQ}$ & P1 & $\sigma_{SEQ}/\sigma_{SWQ}$ & P1 & $\sigma_{SEQ}/\sigma_{NEQ}$ & P1 \\
\hline
 WILC7 &4.6  &0.7\% &4.6 &0.7\% &6.7 &0.2\%\\ 
 WILC5 &4.2 &1.3\% &4.1 &1.3\% &7.0 &0.2\%\\ 
 WILC3 &3.9 &1.5\% &4.1 &1.3\% &6.8 &0.2\%\\
 
   \hline\hline
 Map & $\sigma_{SEQ}/\sigma_{NWQ}$ & P2 & $\sigma_{SEQ}/\sigma_{SWQ}$ & P2 & $\sigma_{SEQ}/\sigma_{NEQ}$ & P2 \\
 \hline
 WILC7 &4.6 &1.7\% &4.6 &1.7\% &6.7 &0.4\%\\ 
 WILC5 &4.2 &2.6\% &4.1 &2.9\% &7.0 &0.3\%\\ 
 WILC3 &3.9 &3.9\% &4.1 &2.9\% &6.8 &0.4\%\\
\hline
\end{tabular}
\end{center}
\end{table}

\end{document}